\documentclass[aps,dvipsnames,twocolumn,amssymb,floatfix,preprintnumbers,amsmath,amssymb,superscriptaddress]{revtex4}
\usepackage{graphicx}%
\relax

\begin{document}

\title{Granular superconductivity below 5~K in SPI-II pyrolytic graphite}
\author{Ana Ballestar}
\affiliation{Division of Superconductivity and Magnetism, Institut
f\"ur Experimentelle Physik II, Universit\"{a}t Leipzig,
Linn\'{e}stra{\ss}e 5, D-04103 Leipzig, Germany}
\author{Pablo Esquinazi}\email{esquin@physik.uni-leipzig.de}
\affiliation{Division of Superconductivity and Magnetism, Institut
f\"ur Experimentelle Physik II, Universit\"{a}t Leipzig,
Linn\'{e}stra{\ss}e 5, D-04103 Leipzig, Germany}
\author{Winfried B\"ohlmann}
\affiliation{Division of Superconductivity and Magnetism, Institut
f\"ur Experimentelle Physik II, Universit\"{a}t Leipzig,
Linn\'{e}stra{\ss}e 5, D-04103 Leipzig, Germany}

\begin{abstract}
We have studied the transport properties of transmission electron microscope (TEM) lamellae obtained from
a pyrolytic graphite sample of grade B (SPI-II) with
electrical contacts at the edges
of the graphene layers. The temperature, magnetic field, input
current dependence of the
resistance as well as the current-voltage characteristic curves
are compatible with the existence of granular superconductivity
below 5~K.  TEM pictures of the studied lamellae
reveal clear differences of the embedded interfaces
to those existing
in more ordered pyrolytic samples, which appear to be the
origin for the
relatively lower temperatures at which the granular superconductivity is observed.
\end{abstract}

\maketitle
\bigskip

\section{Introduction}

Electrical conductivity measurements are an excellent tool to
search for superconductivity embedded in non-superconducting
matrices as in granular superconductors \cite{sha83,deu06} or the
superconductivity located at certain interfaces formed between
materials that are not superconducting  \cite{rey07,goz08}. In
general, it is known that the properties of internal interfaces in
solids can be fundamentally different from those of the
corresponding bulk materials. In graphite samples, either in
Highly Oriented Pyrolytic Graphite (HOPG) \cite{bar08} or in Kish
graphite \cite{ina00}, there exist embedded interfaces. These
two-dimensional (2D) interfaces represent the border between two
graphite crystalline structures with Bernal stacking order but
twisted by an angle $\theta_{\rm twist}$ \cite{esqarx14}.   These
2D interfaces were also found at the surfaces of HOPG samples
\cite{kuw90,war09,flo13,yin14}  as well as in twisted bilayers
graphene,  where several STM studies with $\theta_{\rm twist}
\lesssim 10^{\circ}$ \cite{bri12,gon14} indicate the existence of
van Hove singularities,  in agreement with theoretical estimates
\cite{bis11Mc}.
 We further note that interfaces in pure  and
  doped Bi bicrystals \cite{mun06,mun07} can show superconductivity up to $T_c
\simeq 21~$K \cite{mun08}, although Bi as bulk is not a
superconductor.

Let us clarify to some extent the expected phenomena at the
interfaces, following the discussion in
Ref.~\onlinecite{esqarx14}. A rotation with respect to the
$c-$axis between single crystalline domains of Bernal graphite can
be characterized by a twist angle $\theta_{\rm twist}$ that may
vary from $\sim 1^\circ$ to $< 60^\circ$ \cite{war09}. For twist
angles $> 1^\circ$  the graphene layers (or  graphite sheets)
inside graphite remain unrelaxed giving rise to Moir\'e patterns
\cite{kuw90,war09,bis11,bri12,flo13,yin14}. The experimentally
confirmed enhancement of the density of states at certain regions
of the interface between twisted graphene layers or graphite
sheets \cite{kuw90,war09,bri12,gon14} already suggests a possible
enhancement of the probability to have superconductivity, within
the mean-field BCS general equations.

An  extraordinary result  comes out when the twist angle is small
enough. In this case the graphene or graphite sheets at the
interface can relax their lattices, having perfectly matched
regions of certain size $L(\theta_{\rm twist})$ separated by a
network of screw dislocations. For bilayer graphene with slightly
twisted layers, these networks can be found in Refs.
\onlinecite{ald13,San-Jose2013,gon14}. In this case, i.e. at small
twist angles,  the probability of having high temperature
superconductivity at the interface can be strongly enhanced
because in these dislocation regions the dispersion relation at
low energies becomes flat \cite{San-Jose2013,esqarx14}. The
existence of a flat band in certain regions of the graphite
interfaces is of importance because the relationship between the
critical temperature $T_c$ and the coupling strength between
electrons $g$ changes. Instead of having an exponential
dependence, i.e. $T_c \sim T^\star \exp{-(1/gN)}$ (here $T^\star,
N$ are: a characteristic temperature range where the Cooper pair
coupling applies and the density of states, respectively)  it has
a linear one, i.e. $T_c \sim g V_{FB}$ ($V_{FB}$ is the flat band
volume) \cite{hei11,kop13,kho90,kop11,vol13,mun13,vol14}. We note
that flat bands were found in topological line defects in graphite
\cite{fen12}. Flat bands are believed to be the origin for the
relatively high temperature superconductivity \cite{tan014}
observed at the interfaces of two-layer semiconducting
heterostructures \cite{fog06}.

Moreover, interfaces between rhombohedral (ABCA...) and Bernal
(ABA...) stacking order regions have been observed in graphite
samples \cite{lin12,hat13}, which according to theoretical work
can be  a source for high-temperature superconductivity
\cite{kop13,mun13}. Calculations indicate that high-temperature
superconductivity at the interface can survive throughout the bulk
due to the proximity effect between ABC/ABA interfaces where the
order parameter is enhanced \cite{mun13}. Finally, we should also
note that a second system of dislocations, in this case edge
dislocations, is expected to occur at the boundaries of
crystalline Bernal regions when the tilt angle respect to the
$c-$axis $\theta_{\rm c}\neq 0$.

Experimental evidence for the existence of granular
high-temperature superconductivity in some of those 2D interfaces
has been recently published using HOPG samples of high grade
\cite{bal13,bal14I}, see also \cite{esqpip} and Refs. therein. The
huge field anisotropy in the magnetoresistance \cite{bal13}
indicates that 2D superconductivity should be located at the
interfaces parallel to the graphite sheets.

In this work we are interested on the transport properties of
embedded interfaces in HOPG samples of less order, i.e. grade B,
because in those samples the 2D interfaces either do not exist or
are less clearly defined. As the critical temperature appears to
depend on the size or area of the interfaces \cite{bal14I}, it is
of importance to check whether signs of granular superconductivity
can be found in those less ordered samples, especially at lower
temperatures in comparison to higher quality samples \cite{bal13}.

\section{Experimental details}

For this study we used a HOPG bulk sample of grade II supplied by
SPI, namely SPI-II. This kind of graphite presents a rocking curve
width $\sim 0.8^\circ$ (also known as ZYB grade HOPG). Its purity
has been thoroughly characterized for several but light elements
in Ref.~\onlinecite{spe14} and the results showed that no relevant
concentration of foreign elements exists. In order to check for
the existence of embedded granular superconductivity in the SPI-II
sample we have prepared a transmission electron microscopy (TEM)
lamella  for transport measurements. To prepare the TEM lamella we
used the same procedure as in Refs.~\cite{bal13,bal14I}. The TEM
samples have the advantage that through the contact at the
graphite sheets and interface edges one enhances strongly the
possibility of measuring voltage signals  related to those
interfaces. The usual method of placing the electrical contacts on
the graphite top sheet or at the edges of large samples has the
disadvantage that only a small part of the input current goes
through the interfaces. Consequently,  part of the measured
voltage drop is due to the semiconducting graphene layers
\cite{gar12} and/or the contribution of the $c-$axis resistivity
of graphite with the short circuits one has at the grain
boundaries. The measured TEM lamella had a length $\times$ width
$\times$ thickness equal to $ 20 \times  5 \times 0.5~\mu$m$^3$
and the current and voltage contacts covered the whole
 sample width.  All the measurements have been performed with the four-probe
 electrode technique in the conventional configuration.

\section{Results and discussion}

\begin{figure}[tbp]
\centering
\includegraphics[width=1.0\columnwidth]{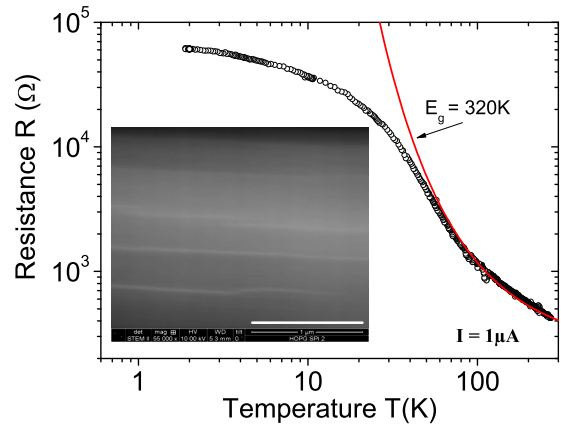}
\caption{Resistance vs. temperature  of the TEM lamella at a fixed
current of $1~\mu$A. The continuous line follows a simple
semiconducting-like exponential function, i.e. $R(T) =   238
[\Omega] \exp{(E_g/2T)}$, where $E_g = 320~$K and $T$ is the
temperature. The inset shows a TEM picture of the measured
lamella. The scale bar at the bottom right represents $1~\mu$m.
The crystalline domains and the borders between them, can be
recognized in this TEM picture by the different gray colors. Note
that the electron beam is applied parallel to the graphite sheets.
The interfaces between regions with a twist angle $\theta_{\rm
twist} > 0$ are located at the borders of regions with uniform
gray colour. Note that no sharp 2D interfaces can be recognized in
this TEM picture, in clear contrast to similar TEM measurements in
highly oriented graphite samples \cite{bar08,bal13}. } \label{RT}
\end{figure}

In the inset of Fig.~\ref{RT} we present a TEM picture of the
studied lamella. The white lines crossing the lamella from left to
right, approximately parallel to the graphene planes (the $c-$axis
is normal to those lines), represent the regions in which the
graphene planes are twisted with respect to the grey-color larger
regions. By comparison with the TEM pictures taken in HOPG grade A
samples, we note that the density of interfaces is much smaller
and they are not as sharp and well defined, see
Refs.~\onlinecite{bar08,esqarx14}.

\begin{figure}[tbp]\centering
\includegraphics[width=1.\columnwidth]{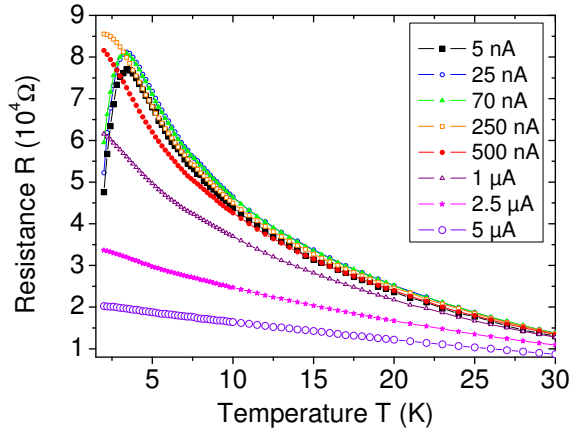}
\caption{Resistance vs. temperature of the TEM lamella at
different  input currents. For clarity we only show the data below
30~K, temperature region where the non-ohmic behaviour is clearly
observed.} \label{RTI}
\end{figure}

Figure~\ref{RT} shows the temperature dependence of the resistance
measured at a constant input current of 1~$\mu$A. Starting from
300~K, the lamella shows a semiconducting-like behaviour down to
$T \sim 70~$K. Below that temperature the resistance increases
slower and tends to saturate at the lowest measured temperatures.
To check for a possible non-ohmic behavior of the resistance, we
performed measurements at different input currents. These results
are shown in Fig.~\ref{RTI} for temperatures below 30~K. The
results at $T \gg 30~$K are current independent (not shown here
for clarity). We observe a clear overall increase of the
resistance decreasing the input current down to 250~nA. However,
for smaller input currents the resistance develops a clear maximum
at $T \sim 3~$K, see Fig.~\ref{RTI}. The further results we show
below indicate that the observed non-ohmic behaviour is related to
the existence of superconducting grains embedded in a
semiconducting matrix.

\begin{figure}[tbp]\centering
\includegraphics[width=0.95\columnwidth]{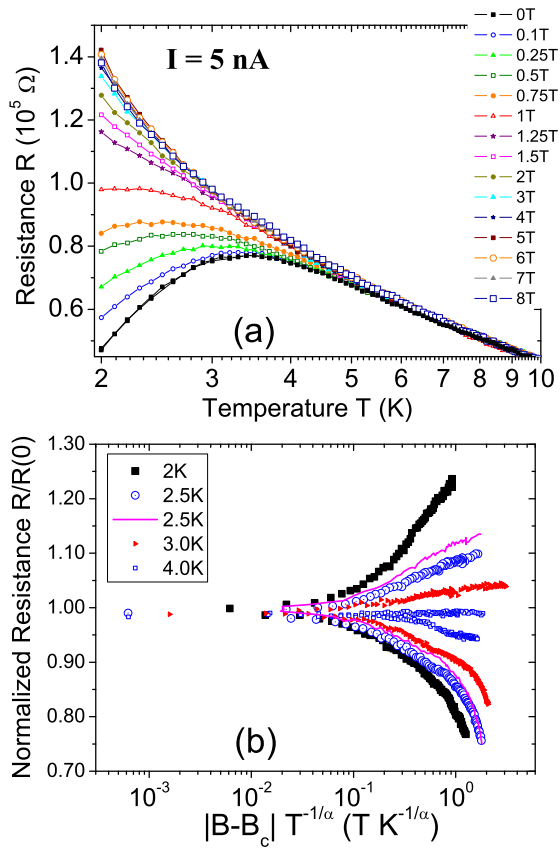}
\caption{ (a) Resistance vs. temperature (semilogarithmic scale)
at different applied magnetic fields (0~T bottom curve to 8~T
upper curve (open squares symbols)) and at an input current of
5~nA. (b) Data obtained from the magnetoresistance measured at
5~nA input current and at different temperatures (not shown here).
The $x-$axis is calculated as the absolute field difference $|B -
B_c|$ multiplied by the corresponding temperature elevated to the
exponent $-1/\alpha$. The curves given by the symbols are obtained
at different fixed temperatures and a critical field $B_c = 1$~T
and $\alpha = 0.7$. The continuous (red) line at 2.5~K was
obtained decreasing the critical field to $B_c = 0.95$~T. The
normalization factor in the $y-$axis is $R(|B-B_c| = 0) =
100,300~\Omega$ at all temperatures.} \label{sca}
\end{figure}

The developed maximum in the resistance vs.  temperature at low
input currents already suggests that a coupling between the
superconducting grains could exist. An important hint for the
existence of granular superconductivity  is provided by the
 magnetic field dependence of the resistance,  especially below its maximum.
 Figure~\ref{sca}(a) shows the resistance vs.  temperature at different  magnetic fields
 applied normal to the graphene planes.
 Those measurements have been done at constant input current of   5~nA. The observed
 behaviour  resembles  the one seen in the resistance of bulk HOPG samples
 of similar grade but with contacts at the graphene top layer, i.e. at the sample
 surface \cite{yakovssp,kempa00}.
 We further note that this behaviour is qualitatively
 identical to the one observed in certain higher grade HOPG
 samples \cite{yakovprl03,yakovadv03,tok04,heb05},
 where the metallic-insulating field-driven change
 is observed up to
 room temperature.
 This magnetic field-driven transition, the
 so-called metal-insulator transition (MIT), was
 discussed in several papers \cite{kempa00,yakovprl03,yakovadv03} in terms
 of superconducting-insulator transition because of its analogy
 to the one observed in MoGe superconducting films \cite{mas99}.  The MIT in graphite has also
 been interpreted using  quasi classical transport equations
 within the 3D electronic  band structure of graphite \cite{mcc64} but with several free
 magnetic-field dependent parameters \cite{tok04,heb05}. We note, however, that the MIT
 is observed only if interfaces in the graphite sample exist.
The metallic-like behavior vanishes for thin enough graphite
samples \cite{bar08,van11,gar12} and obviously the MIT does not
exist anymore, indicating that this transition is not intrinsic of
the
 ideal graphite structure.

 In bulk HOPG samples the observed MIT shows an interesting scaling
 that follows to a certain extent the scaling theory for
 quantum phase transitions in disordered two-dimensional superconductors \cite{fis90}.
In this case the resistance in the critical regime is given by the
equation $R(\delta, T) = R_{c} f(| \delta |, T^{1/\alpha})$ where
$R_{cr}$ is the resistance at the transition, $\delta = B -
B_{c}$, $B_{c}$ is a critical magnetic field, $f$ a scaling
function such that $f(0) = 1$ and $\alpha$ a critical exponent. In
bulk HOPG samples with MIT excellent  scalings have been obtained
with $B_{c} \simeq 0.1 \ldots 0.2~$T and $\alpha = 0.65 \pm 0.05$
\cite{kempa00,yakovadv03}. With the data shown in
Fig.~\ref{sca}(a), using $B_c = 1~$T and $\alpha = 0.7$ we get the
results shown in Fig.~\ref{sca}(b). Although there is a clear
similarity to the behaviour reported in bulk HOPG samples
\cite{kempa00} there is no perfect scaling, i.e. the curves
deviate from each other and look less symmetric respect to the
critical $R/R(0) = 1$ line as the temperate increases. The reason
for this deviation is the input current dependence of the
resistance, see Fig.~\ref{RTI}, a dependence that does not exist
in bulk HOPG samples mainly due to the distribution of the current
along several graphite sheets and interface regions in parallel.
Qualitatively speaking the effect of the current with temperature
in the scaling behaviour of Fig.~\ref{sca}(b) can be partially
compensated decreasing the critical  field with temperature. As
example, in Fig.~\ref{sca}(b) we show the same data at 2.5~K but
with a critical field of 0.95~T instead of 1~T. One recognises the
shift of the upper curve towards the 2~K data and a better
symmetry respect to the critical line.   We note further that for
the TEM lamella we get a critical exponent $\alpha$ similar to the
one obtained in bulk HOPG samples, however the critical field is
one order of magnitude higher \cite{kempa00,yakovadv03}. This
appears to be related to some characteristic size of the
superconducting regions, as we will discuss below.

\begin{figure}[tbp]\centering
\includegraphics[width=1.0\columnwidth]{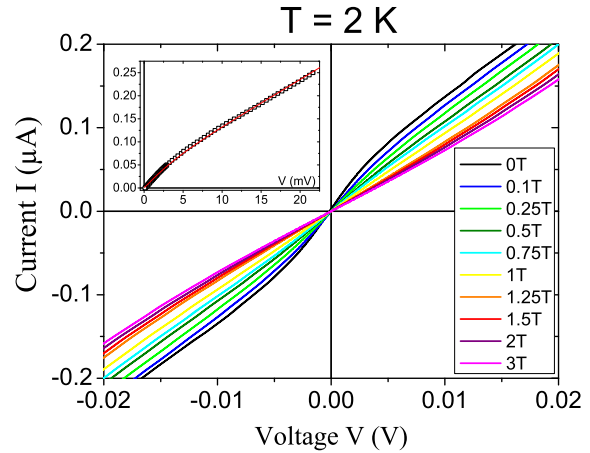}
\caption{Current-voltage characteristic curves obtained for the
TEM lamella at different applied fields at  2~K. The inset shows
the curve at zero field with a fit (continuous line) to the
differential equation of Ref.~\cite{amb69}, using as a free
parameter the Josephson critical current $I_c = 82~$nA.}
\label{iv2k}
\end{figure}

\begin{figure}[tbp]\centering
\includegraphics[width=0.95\columnwidth]{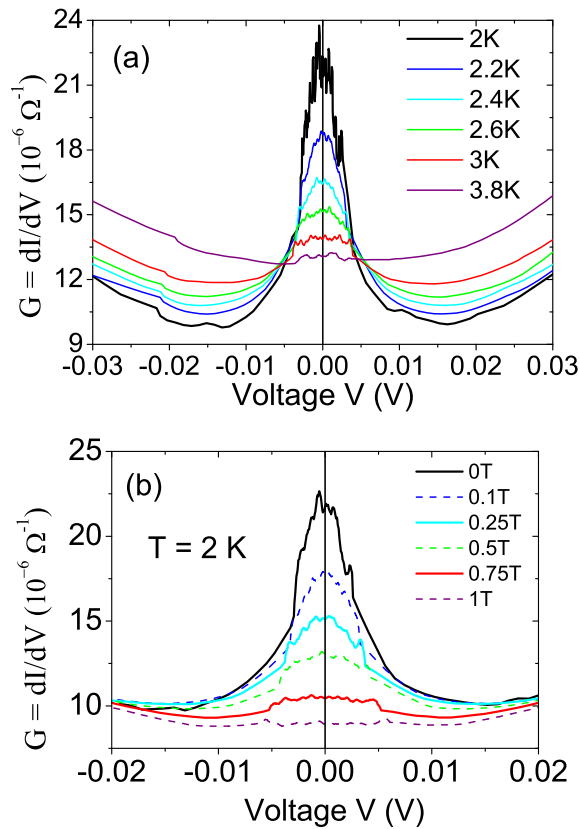}
\caption{(a) Differential conductivity measured at temperatures
between 2~K and 3.8~K, i.e. in the temperature range where a
finite coupling between the superconducting regions exists. (b)
Differential conductivity at 2~K measured at different magnetic
fields. The zero field data are the same as in (a) after an
average smoothing to remove the noise at low voltages. Above 1~T
the non-linear effect associated to superconducting behavior
clearly vanishes.} \label{g}
\end{figure}

The current-voltage
$(I-V)$ characteristic curves at fixed temperature suggest the existence of
 the Josephson effect between  superconducting grains
embedded in our lamella. Figure \ref{iv2k} shows the $I-V$ curves
obtained at 2~K and at different applied magnetic fields. The
curve at zero field as well as the changes observed with field
appear compatible with the existence of the Josephson effect.
However, even  the curve at zero field does not show zero
resistance at any current, within experimental resolution. This
can be due to, either  a finite resistance in series to the
embedded Josephson coupled regions, to thermal fluctuation
effects, or to both effects simultaneously. The voltage dependence
of the differential conductivity $G=\rm{d}I/\rm{d}V$ at zero field
and different temperatures or at a fixed temperature and different
applied fields, see Fig.~\ref{g}, is compatible with the existence
of granular superconductivity. The temperature dependence of the
conductivity peak at zero voltage, see Fig.~\ref{g}(a), can be
tentatively used  to estimate  by extrapolation the  conductivity
at zero temperature. In the, obviously restricted, temperature
range the peak in the conductivity follows very well a thermally
activated exponential function $G[\Omega^{-1}] \simeq 1.94 \times
10^{-3} \exp{(-T/0.38)} + 1.3 \times 10{-5}$, as commonly observed
in granular superconductors \cite{sha83}. The obtained prefactor
of the exponential function indicates that the conductivity would
grow up to values orders of magnitude larger than the conductivity
at high temperatures. Therefore, we assume that the finite
conductivity at 2~K is mainly due to thermal fluctuations and not
due to an extra resistance in series.

As it was done in previous studies \cite{bal13,bal14I} we account
for the thermal fluctuation effects  in a Josephson junction (the
thermal energy $k_B T$ is larger than the Josephson coupling
energy $E_J$ = $(\hbar/2e)I_c$, $I_c$ is the critical Josephson
current) using  the differential equation proposed in
\cite{amb69,iva68}. We fit the measured $I-V$ curves with this
model (with the critical current $I_c(T)$ as the only free
parameter) and we obtain a very good agreement, as shown in the
inset of Fig.~\ref{iv2k}.

Finally, let us discuss the possible reason for the,
comparatively, low  temperature at which the Josephson behaviour
sets in. We can estimate the size of the Josephson junction using
the critical field $B_c \sim 1~$T obtained from the resistance
measurements, see Figs.~\ref{sca} and \ref{iv2k}. Following the
same arguments as in Ref.~\cite{bal14I} we assume that the
critical field $B_c$ would produce a single flux quantum in an
effective  Josephson area. In this case we estimate an area of the
order of $45 \times 45~$nm$^2$. This small size and the low
critical temperature are compatible with  the study on the size
dependence of the critical temperature of lamellae with interfaces
done in Ref.~\cite{bal14I}. Although the origin of this size
dependence is not yet completely clarified, we note that this size
dependence was already found in conventional superconducting
multilayers and thin wires. It has been tentatively interpreted
\cite{gui86,OF} as an phenomenon related to weak localization
 corrections to $T_c$ for 2D superconductors \cite{fuk84}.
The key factor to understand the results is the presence of
disorder that affects the screening of the Coulomb interaction and
therefore the BCS coupling parameter, resulting in an exponential
suppression of the critical temperature.

The experimental evidence for the existence of interfaces, the
enhancement of the density of states in some of those interfaces
regions as well as the possible existence of anomalously flat
bands, provide  a strong support to recent theories on the
importance of flat bands to trigger 2D high temperature
superconductivity. Nevertheless, till all the necessary
experimental evidence is obtained,  we should not rule out other
possibilities that could explain the observed behavior,  for
example, charge density waves (CDW). We note that evidence for the
formation of CDW was found in CaC$_6$ at $\sim 250~$K, a material
that becomes superconducting at $T_c = 11.5~$K \cite{rah11}.
Although CDW is antagonist to superconductivity \cite{gab13}, some
coexistence with superconductivity has been reported in systems
like NbSe$_2$ \cite{sou13} or EuBiS$_2$F \cite{zha14}. The
question arises, whether a kind of CDW at the interfaces could be
part of the origin of the observed behavior.

\section{Conclusion}

In summary, the obtained results together with the characteristics
of the interfaces observed in TEM pictures for SPI-II HOPG
samples, suggest that the superconducting regions at the
interfaces are of smaller size than in more ordered HOPG samples,
i.e. HOPG grade A. Moreover, the disorder may play a relevant role
in lowering the critical temperature where
 granular superconductivity is observed. The obtained behavior
 with magnetic field and temperature resembles the
 metal-insulator transition (MIT) found in graphite
 \cite{kempa00,yakovprl03,tok04,heb05}.
 Our results
indicate that the MIT can be indeed related to superconductivity.
The overall results support the existence of superconductivity
located at interface regions. Upon the characteristics of the HOPG
sample,  superconductivity can occur at very high temperatures.


This work was partially supported by the Europ\"aischen
Sozialfonds (ESF) 100124929.



\end{document}